\documentclass[twocolumn]{article}
\usepackage{icqproc,amsmath}
\usepackage{psfig}

\begin{document}
\thispagestyle{empty}
\twocolumn[
\vspace*{30mm}
\begin{LARGE}
\begin{center}
Tile Hamiltonians for Decagonal Phases
\end{center}
\end{LARGE}
\begin{large}
\begin{center}
Michael Widom,
Ibrahim Al-Lehyani and Marek Mihalkovic
%
%  END AUTHORS
%
\end{center}
\end{large}
\begin{footnotesize}
\begin{it}
\begin{center}
%
%  ADDRESS
%
Department of Physics, Carnegie Mellon University, Pittsburgh, PA  15213 USA
%
%  END ADDRESS
%
\end{center}
\end{it}
\end{footnotesize}

\begin{footnotesize}
\begin{center}
%
%  DATE
%
\today
%
%  END DATE
%
\end{center}
\end{footnotesize}

\vspace{4ex}
\begin{small}
\hrule\vspace{3ex}
\begin{minipage}{\textwidth}
{\bf Abstract}\vspace{2ex}\\
\hp 
%
%  ABSTRACT
%
A tile Hamiltonian (TH) replaces the actual atomic interactions in a
quasicrystal with effective interactions between and within tiles. We
study Al-Co-Ni and Al-Co-Cu decagonal quasicrystals described as
decorated Hexagon-Boat-Star (HBS) tiles using {\em ab-initio}
methods. A dominant term in the TH counts the number of H, B and S
tiles, favoring tilings of H and B only. In our model for Al-Co-Cu,
chemical ordering of Cu and Co along tile edges defines tile edge
arrowing.  Unlike the edge arrowing of Penrose matching rules,
however, the energetics for Al-Co-Cu do not force quasiperiodicity.
Energetically favored structures resemble crystalline approximants to
which the actual quasicrystalline compounds transform at low
temperature.
%
%  END ABSTRACT
%
\vspace{2.5ex}\\
\end{minipage}\vspace{3ex}
\hrule
\end{small}\vspace{6ex}
]

%
%  MAIN TEXT
%

Explaining thermodynamic stability is a fundamental problem in the
field of quasicrystals. Competing explanations range from {\em
energetic stabilization} utilizing matching rules such as those that
force quasiperiodicity in the Penrose tiling~\cite{Ingersent}, to {\em
entropic stabilization}~\cite{WSS} focusing on the configurational
entropy available in random tiling models~\cite{elser,henley}.
Experimental evidence so far has not unambiguously settled the matter,
and the true situation is certainly more complex than either of the
two extremes just described. With the advent of plausible atomistic
quasicrystal models and advances in first-principles calculation
methodology we hope further theoretical progress may be made in this
area.

Our approach reported here is based on a ``tiling Hamiltonian'', in
which a family of low energy atomistic structures is placed in 1:1
correspondence with a family of tilings of the plane. The energetics
of the tiling Hamiltonian is defined in a manner that closely
approximates the {\it ab-initio} energetics of the atomistic
structures. The energetics we derive proves remeniscient of Penrose
``matching rules'' (which force global quasiperiodicity in
minimum-energy structures) but differs in crucial aspects. Indeed, we
find that our tile Hamiltonian does {\it not} favor quasiperiodicity.
Quasiperiodicity may occur at high temperatures as a result of random
tiling configurational entropy. At low temperatures energy favors
transformation to crystalline phases, which is indeed often observed
experimentally~\cite{ritsch8,Hiraga_AlCoCu,Fettweis,Hiraga_AlCoNi,Doblinger}.

\begin{figure}[tbh]
\psfig{file=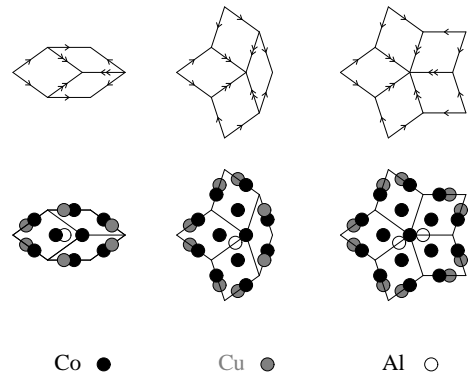,width=3in}
\caption{(A) HBS tiles and their decompositions to Penrose tiles.
(b) Atomic decorations for AlCoCu. In (b), only TM and symmetry
breaking Al atoms are shown. For AlCoNi both edge sites are decorated
with Ni atoms.}
\label{fig:penrose}
\end{figure}

Penrose tiles are fat and thin rhombi (Fig~\ref{fig:penrose}a). Edges
are assigned single- and double-arrow decorations (as shown) which
must match for common edges in adjacent tiles. Perfect quasicrystals
obey these rules everywhere. The double-arrow matching
rule~\cite{henley} causes rhombi to associate into hexagon (H), boat
(B) and star (S) shapes (with relative frequency
H:B:S=$\sqrt{5}\tau$:$\sqrt{5}$:1), while the single-arrow rules force
quasiperiodicity in the HBS tiling.  It has been shown
previously~\cite{CWmodel,au8} that plausible atomistic structures of
AlCoNi and AlCoCu can be described as HBS tilings decorated with atoms
(Fig~\ref{fig:penrose}b). Hence, we may consider the Penrose rhombus
double-arrow rules to be satisfied by definition of our basic HBS
tiles.

In a tiling model of quasicrystals, the actual atomic interactions in
the system Hamiltonian can be replaced with effective interactions
between and within tiles~\cite{tileham}. The resulting tile
Hamiltonian is a rearrangement of contributions to the actual total
energy. In a simple atomic interaction picture (pair potentials for
example) the relation between the actual atomic interactions and the
tile Hamiltonian is straightforward. It might be difficult to find the
relations between them for more complicated atomic interactions (many
body potentials, or full {\em ab-initio} energetics, for example) but
it is theoretically possible. The tile Hamiltonian includes terms
which depend only on the number of tiles, and includes other terms for
tile interactions. The tile Hamiltonian greatly simplifies our
understanding of the relationship between structure and energy, and is
a reasonable way to describe the tiling ensemble.

\begin{figure}[tbh]
\psfig{figure=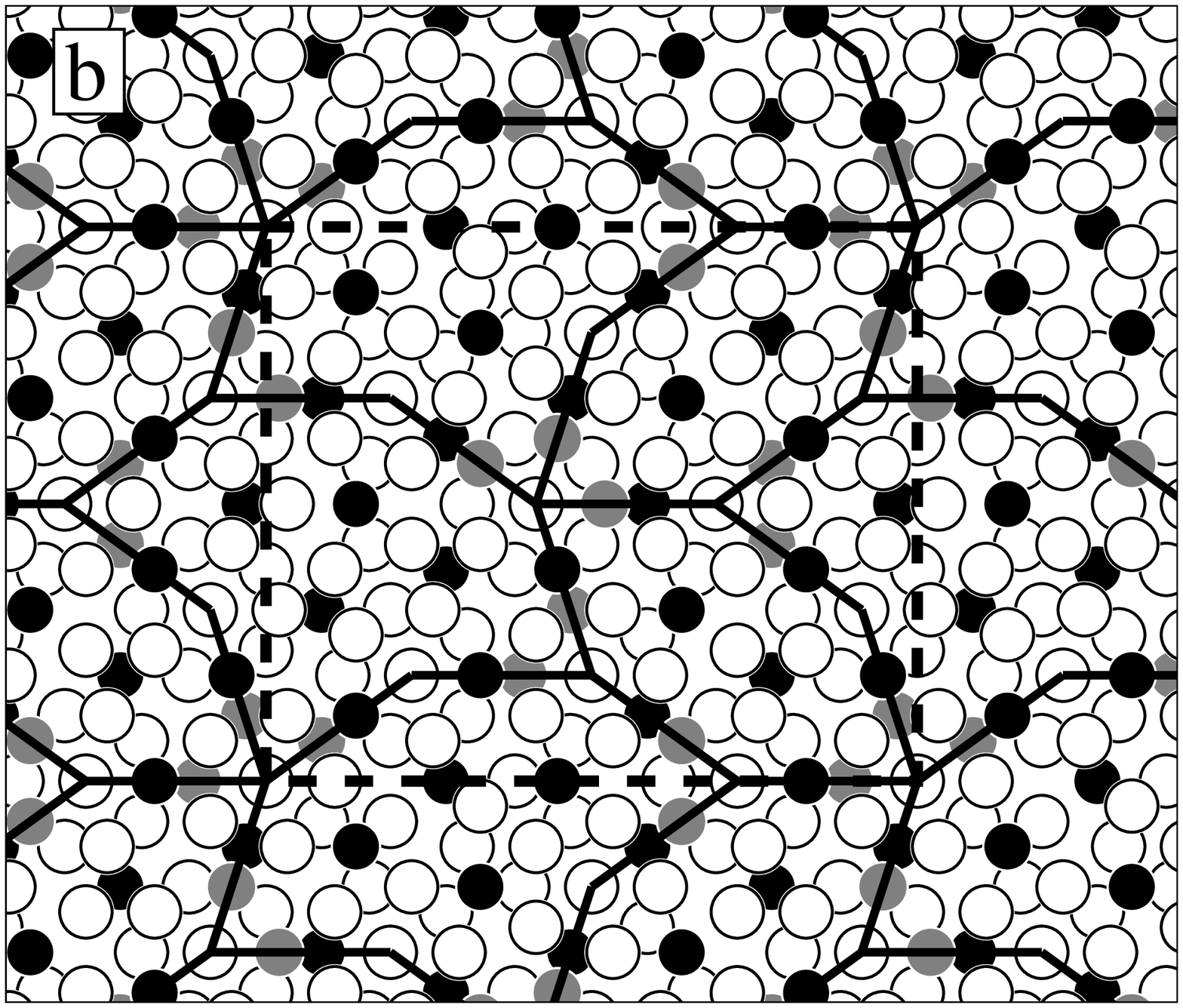,width=3in,angle=-90}
\psfig{figure=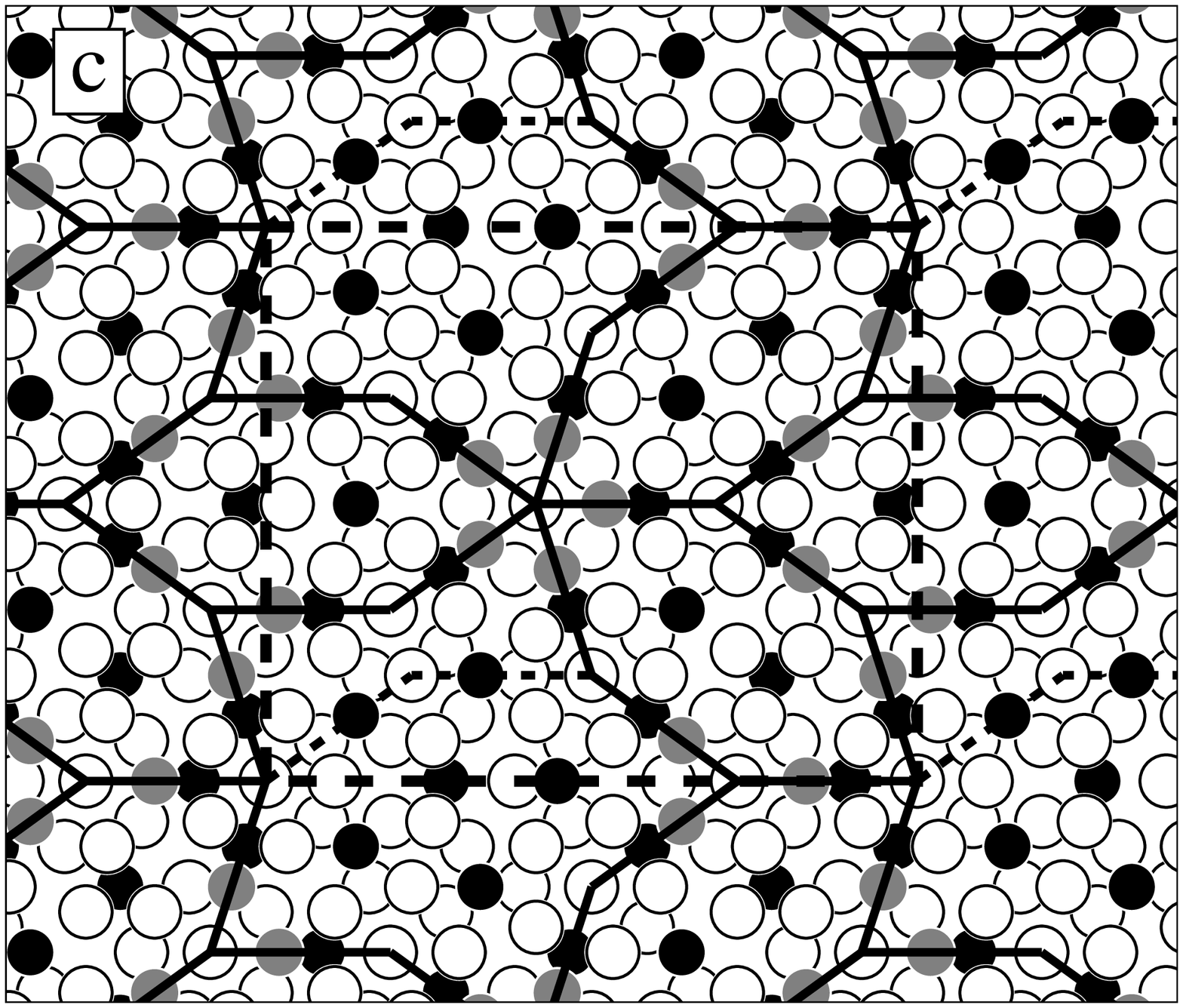,width=3in,angle=-90}
\caption{Space can be tiled in many ways using HBS tiles. Both these 
approximants contain equal numbers of each atom type. Structures in
(a) and (b) differ by the phason flip outlined in (b) with a dashed
line, which converts two B tiles into and H and an S.}
\label{fig:app132}
\end{figure}

Are Penrose single-arrow matching rules enforced by energetics of real
materials? For a simple model~\cite{au8} of Al$_{70}$Co$_{9}$Ni$_{21}$
in which both edge sites are occupied by Ni atoms there is no source
of symmetry-breaking at short length scales able to define an
orientation of the tile edges. The energetics of structures based on
HBS tiles decorated in this manner depends primarily on the numbers of
H, B and S tiles. As seen in Fig.~\ref{fig:app132}, certain phason
flips convert an HS pair into a BB pair (or
vice-versa). Pair-potential-based total energy calculations of these
two structures~\cite{au8,sendai} reveal that structure (a) containing
the BB pair is lower in energy than (b) containing the HS pair by 0.2
eV. The physical origin of this energy difference lies in the number
of 72$^{\circ}$ vertices, which drops by 1 in the transition
HS~$\rightarrow$~BB. At a 72$^{\circ}$ vertex transition metal pairs
are close neighbors, causing a reduction in the number of
energetically favorable~\cite{corrections} aluminum-transition metal
near neighbor interactions.

Hence we may express the tile Hamiltonian as
\begin{equation}
H=E_s N_s
\label{eq1}
\end{equation}
where $N_s$ is the number of star tiles present, and the coefficient
$E_s = 0.2$ eV. To fully model decagonal AlCoNi (indeed any decagonal
phase) we should add into the Hamiltonian (eq.~(\ref{eq1})) terms
representing phason stacking disorder. Unfortunately, at this time the
magnitude (and even the sign) of this term is
unknown~\cite{sendai}. Additional corrections relating to the number
of 144$^{\circ}$ vertices are small relative to the term
shown~\cite{au8,sendai}. Hence we focus our attention on the
two-dimensional behavior defined by Hamiltonian (eq.~(\ref{eq1})).

\begin{figure}[tbh]
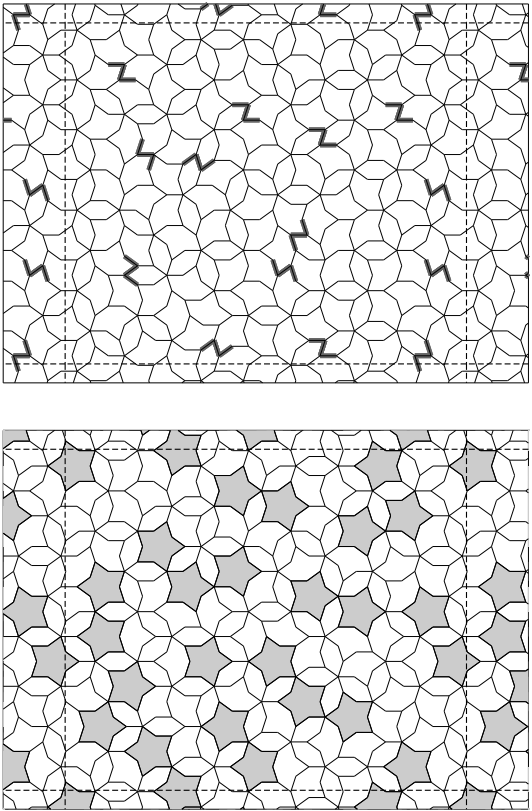

\psfig{figure=521-HBZZ.ps,angle=90,width=3in}
\psfig{figure=521-HBS.ps,angle=90,width=3in}
\caption{Typical low temperature configurations. (a) Tile Hamiltonian
(eq.~(\ref{eq1})), or (eq.~(\ref{eq2})) with $E_s>2E_{zz}$. Wide gray
bonds identify ``zig-zag'' structures. (b) Tile Hamiltonian
(eq.~(\ref{eq2})) with $E_s<2E_{zz}$. Star tiles are shaded gray for emphasis.}
\label{lowT}
\end{figure}

Monte Carlo simulations show that S tiles are infrequent at $T=1000$K and
completely absent in the lowest energy structures, which are random HB
tilings with relative frequency H:B=1:$\tau$. A typical structure is
illustrated in Fig.~\ref{lowT}a.

The situation for AlCoCu is more complicated than for AlCoNi, due to
the chemical alternation of Co/Cu pairs on tile edges.  Cockayne and
Widom~\cite{CWmodel} suggested that tile edges could be assigned arrow
direction based on their Co/Cu decorations (Fig.\ref{fig:penrose}b).
The physical origin of Co/Cu chemical ordering rests on the status of
Cu as a Noble Metal with completely filled d orbitals, unlike normal
transition metals such as Co. Energetically, it turns out to be highly
favorable for Co/Cu pairs to orient such that the Co atoms are further
removed from 72$^{\circ}$ vertices than Cu atoms.

For consistency with Penrose matching rules, we thus define the arrow
to point from Cu towards Co. When the HBS tiles are decorated
consistently with the Penrose matching rules, all arrows point
outwards from 72$^{\circ}$ vertices, minimizing the energy associated
with chemical ordering of Co/Cu. However other tilings (such as the
random HB tiling illustrated in Fig.~\ref{lowT}a) contain ``zig-zag''
structures. The middle of the three bonds in a zig-zag can never be
oriented to point outwards from each of its 72$^{\circ}$ vertices,
leading to a minimum energy cost for each zig-zag, $E_{zz}$. Hence we
define our tile Hamiltonian
\begin{equation}
H=E_s N_s + E_{zz} N_{zz}
\label{eq2}
\end{equation}
where $N_{zz}$ is the number of zig-zags present, while the
coefficients $E_s = 0.2$ eV and $E_{zz} = 0.12$ eV have been derived
from full {\em ab-initio} calculations~\cite{AW}. Eq.~(\ref{eq2}) is
actually a simplification of the full tile Hamiltonian~\cite{AW,ICQ7}
which captures accurately the energetics of the lowest energy
structures. Although we assign the energy cost to the zig-zag shape,
its origin is the frustration of the central bond orientation, and not
a feature of the shape itself.

\begin{figure}[tbh]
\psfig{figure=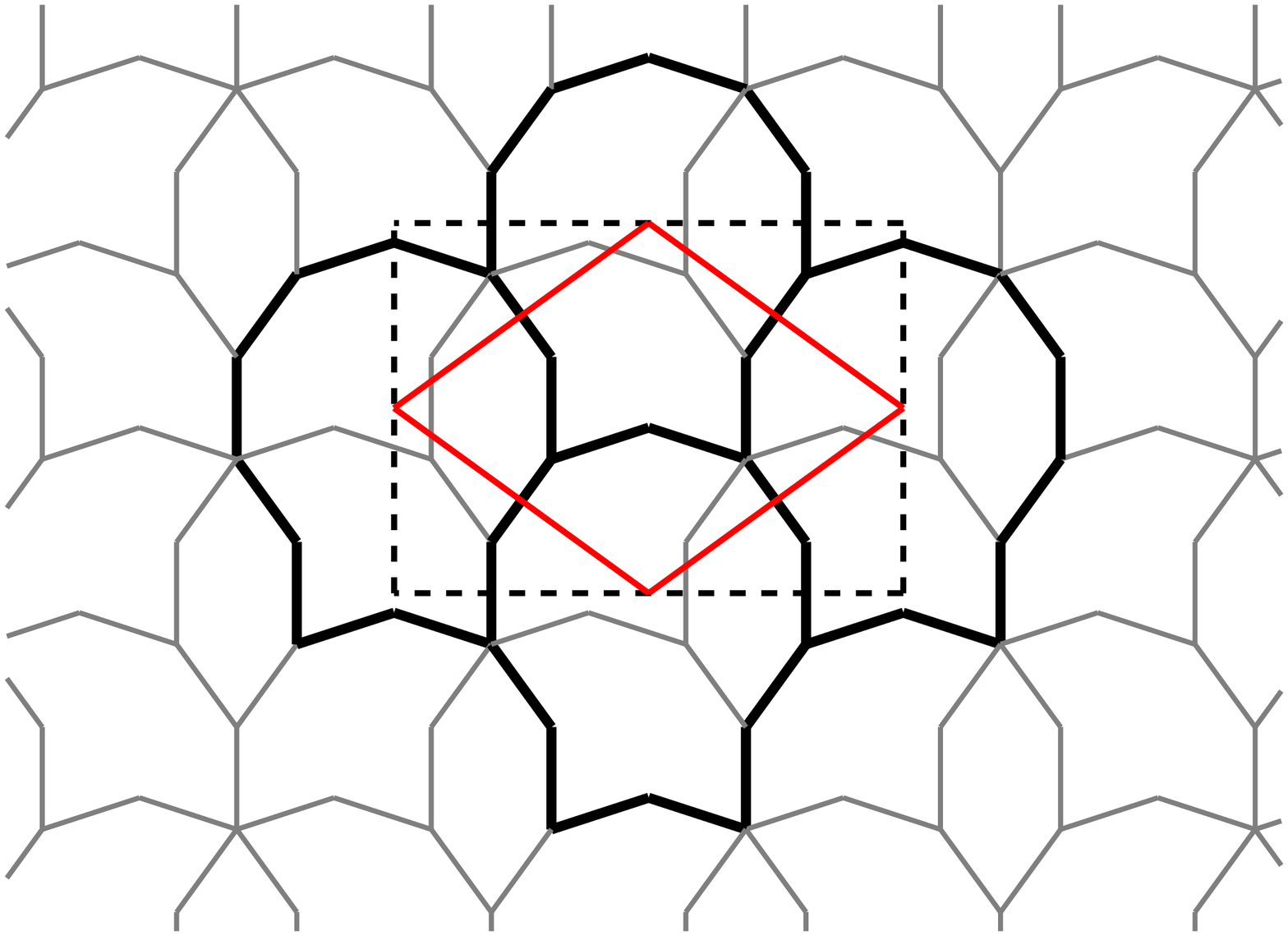,angle=-90,width=3in}
\hspace{0.05in}
\psfig{figure=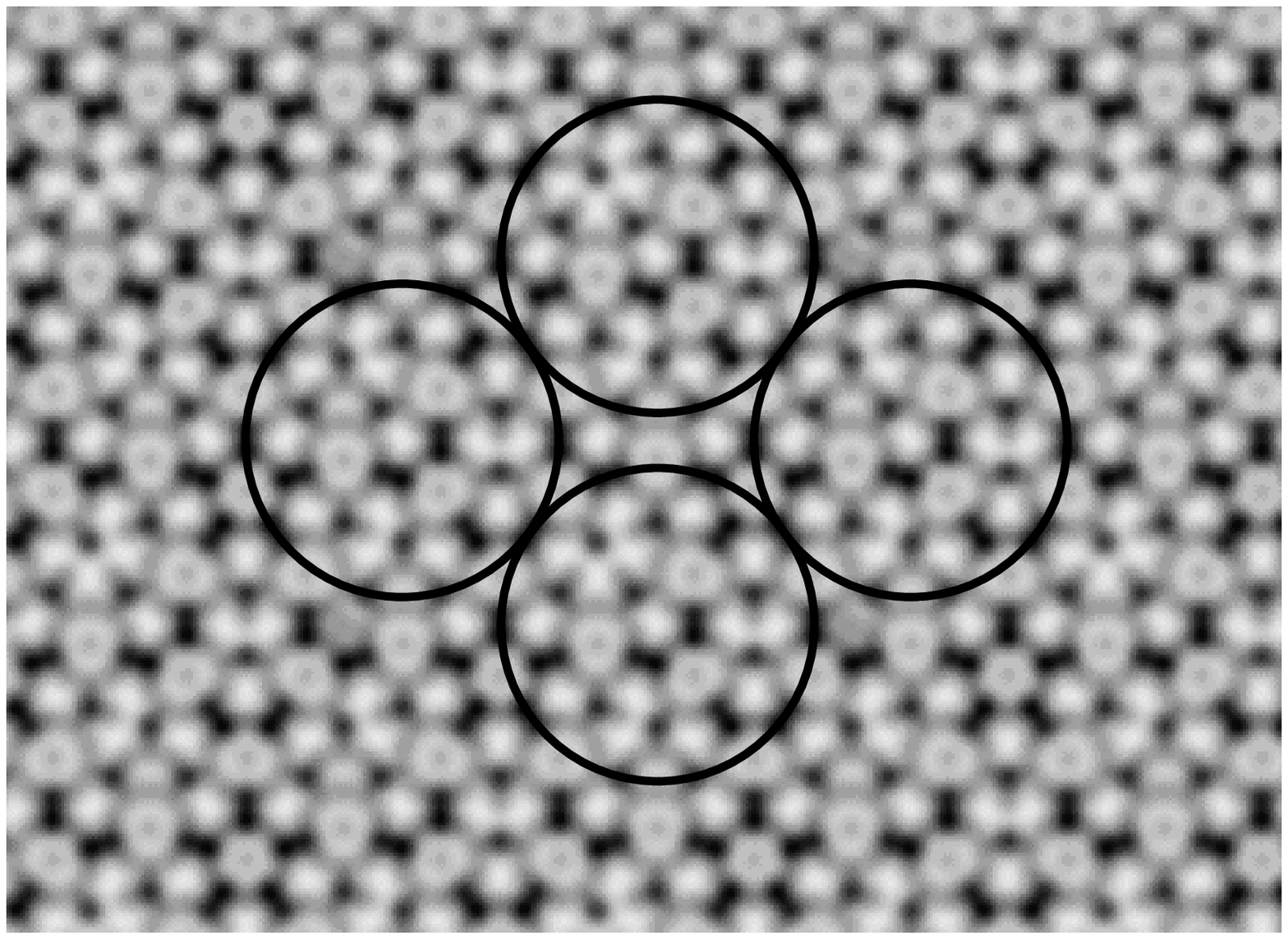,width=2.75in}
\caption{zig-zag- and star-free ``lightbulb tiling'' structure.
(a) Tiling has orthorhombic cell with lattice constants
23.1~\AA~$\times$ 31.8~\AA~ (dashed lines) containing inscribed
72$^{\circ}$ rhombus with 20~\AA~ edge length. (b) Model HREM
structure image~\cite{HREM} showing 20~\AA~ ring contrasts at vertices
of 72$^{\circ}$ rhombus.}
\label{noSZZ}
\end{figure}

We have found a few special approximants containing neither stars nor
zig-zags.  The simplest of these approximants (and the largest phason
strain) covers space by translation of a single boat tile (see tiling
B1 in Ref.~\cite{AW}).  The next larger of these approximants (but the
smaller phason strain) covers space with ``lightbulb'' objects (see
Fig.~\ref{noSZZ}a) consisting of two boats and a hexagon. Other star-
and zig-zag-free structures have been found that are basically
superstructures of the lightbulb tiling.

For large quasicrystal approximants of low phason strain it appears
impossible to simultaneously eliminate both stars and zig-zags. Were
we to start with a phason strain-free random HB tiling, containing $N$
tiles ($N/\tau^2$ tiles of type H and $N\tau$ tiles of type B), a
series of tile flips could segregate the tiles into a zig-zag-free
lightbulb tiling adjoining a pure H tiling. Counting up the tile
numbers, we see that the lightbulb tiling contains $N/\tau^2$ type B
tiles and hence $N/2\tau^2$ type H tiles. This leaves
$(1/\tau-1/2\tau^2)N$ extra H tiles remaining to form a pure H tiling
which contains 1 zig-zag per H tile. Accordingly, we conjecture
$(1/\tau-1/2\tau^2)N$ is the minimum number of zig-zags possible in an
HB tiling of $N$ tiles at composition H$_1$B$_{\tau}$. The number of S
tiles present in an ideal HBS tiling of $N$ tiles total works out to
$(2/\tau-1/\tau^2)N$, just twice the apparent minimum number of
zig-zags. Indeed, we believe this may be the minimal allowed value of
$N_s+2N_{zz}$ in zero phason strain tilings. If this were true, then
the density of stars in a zero phason strain Penrose tiling is the
minimum possible density of stars in any zig-zag-free tiling.

The lightbulb tiling illustrated in Fig.~\ref{noSZZ}a exhibits a unit
cell of a $72^{\circ}$ rhombus with an edge length of
$2(\cos{{\pi}\over{10}}+\cos{{3\pi}\over{10}})L
\approx 20$~\AA~ where $L=6.4$~\AA~ is the edge length of the HBS
tiling for AlCoCu. Such a crystal structure appears when decagonal
Al$_{65}$Co$_{20}$Cu$_{15}$ is annealed at low temperatures. It is
seen in HREM as a rhombic lattice of ring contrasts identified as
$20$~\AA~ clusters. HREM images of the atomic structure associated
with our lightbulb tiling (when decorated with atoms as in
Fig.~\ref{fig:app132}) contain nearly complete ring contrasts.
Fig~\ref{noSZZ}b illustrates a simple model high-resolution structure
image~\cite{HREM} obtained by superposing Gaussian functions at each
atomic position with weight proportional to the atomic number (to do a
better job of HREM modeling we should incorporate chemical and phason
stacking disorder in our structure model and perform dynamical
diffraction analysis of the electron microscope imaging). Dark spots
correspond to atomic columns and white to empty channels. This type of
image should resemble HREM images from a thin sample near the Scherzer
defocus. Fig.~\ref{noSZZ} bears a qualitative resemblence to the HREM
patterns of low-temperature Al$_{65}$Co$_{20}$Cu$_{15}$ in
Ref.~\cite{Hiraga_AlCoCu}.  Thus it may be that our tiling Hamiltonian
gives an indication of the structure of the low temperature phase, and
explains its appearance as driven by the need to eliminate star tiles
and zig-zags.

The precise values of the coefficients $E_s$ and $E_{zz}$ in
equations~(\ref{eq1}) and~(\ref{eq2}) can be questioned because they
wre calculated with atoms placed at ideal sites. Their values will
change if atomic relaxation is allowed, although we expect the general
form of the tile Hamiltonians and the magnitudes and signs of the
terms to be preserved. Small changes in chemical composition can lead
to surprisingly large changes in the tile Hamiltonian by altering the
atomic interactions specifically at those points where the unfavorable
star or zig-zag energies originate. Such an effect could explain why
the low temperature structure observed for
Al$_{63}$Co$_{17.5}$Cu$_{17.5}$Si$_2$ (a 72$^{\circ}$ rhombus with a
51~\AA~ edge length~\cite{Fettweis}) differs from that
found~\cite{Hiraga_AlCoCu} in Al$_{65}$Co$_{20}$Cu$_{15}$. In general,
variation of the tile Hamiltonian parameters can lead to transitions
such as that illustrated in Fig.~\ref{lowT}.

\begin{figure}
\psfig{figure=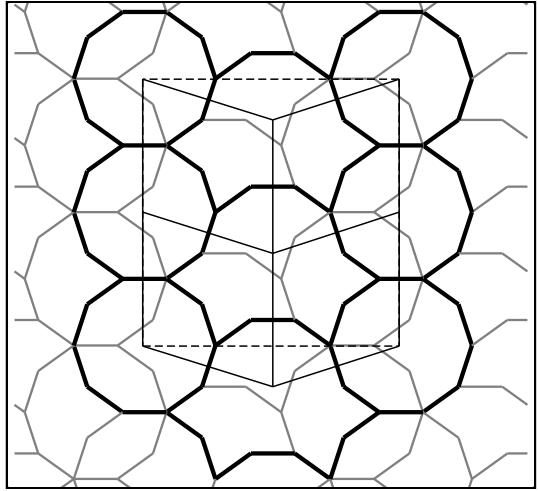,width=2.75in}
\caption{Chevron tiling of 72$^{\circ}$ rhombi with 20~\AA~ edge lengths.
The rhombic unit cell of dimensions 37.6~\AA~$\times$ 39.7~\AA~ matches
the low temperature structure PD1 of Al-Co-Ni~\cite{Doblinger}}.
\label{fig:pd1}
\end{figure}

A chemistry dependence is also found in the case of AlCoNi, where
small changes in composition lead to a wide array of different
structure types~\cite{ritsch8}.  Transitions as composition (or
temperature) is changed may be related to changes in the values of
terms in a tile Hamiltonian. For example, a change from
Al$_{70}$Co$_9$Ni$_{21}$ to Al$_{72}$Co$_{11}$Ni$_{17}$ results in
CoAl pairs replacing NiNi pairs on tile edge sites at 72$^{\circ}$
vertices~\cite{au8}.  Consequently the energy cost of 72$^{\circ}$
vertices, and hence $E_s$ is reduced on average.  However, the Co/Al
pairs carry an edge orientation (similar to Co/Cu pairs) so we need to
add a zig-zag energy into the Hamiltonian~(\ref{eq1}), resulting in a
new Hamiltonian like~(\ref{eq2}). Although the true low temperature
phase at this composition is not certain, at a nearby composition of
Al$_{71}$Co$_{14.5}$Ni$_{14.5}$ the system indeed takes on one of two
structures based on tilings by 72$^{\circ}$ rhombi with 20~\AA~ edge
lengths~\cite{Hiraga_AlCoNi,Doblinger}. One structure, known as PD2,
has the unit cell of the lightbulb tiling (Fig.~\ref{noSZZ}). The
other structure, known as PD1, pairs rhombi into ``chevron''
structures (Fig.~\ref{fig:pd1}) in which, again, both stars and zigzags
may be avoided. Both of these structures have an 8~\AA~ periodicity in
the stacking direction, so an additional term related to phason
stacking faults may need to be included in our tile Hamiltonians.

In conclusion, we show that an ensemble of low energy quasicrystal and
approximant structures may be modeled using very simple tiling
Hamiltonians. The tile Hamiltonians representing Al-Co-Ni and Al-Co-Cu
favor crystalline structures at low temperatures but may exhibit
quasicrystals in equilibrium at high temperatures. The favored low
energy crystal structures resemble the transformation products
actually observed in these compounds at low temperatures.

\section{Acknowledgments}
The authors thank C.L. Henley and S. Naidu for useful discussions. IA
wishes to thank King Abdul Aziz University (Saudi Arabia) for
supporting his study and MW acknowledges support by the National
Science Foundation under grant DMR-0111198. We thank the Pittsburgh
Supercomputer Center for computer time used for this study.

\begin{footnotesize}
\begin{frenchspacing}

\end{frenchspacing}
\end{footnotesize}

\end{document}